\documentclass[aps,prd,twocolumn,floatfix,nofootinbib,preprintnumbers,superscriptaddress]{revtex4-1}

\usepackage[utf8]{inputenc}
\usepackage{newtxtext}
\usepackage[libertine]{newtxmath}

\usepackage{graphicx}
\usepackage{slashed}
\usepackage{amsfonts}
\usepackage{changes}

\usepackage[english]{babel}

\usepackage{blindtext}
\usepackage{microtype}

\usepackage[colorlinks,linkcolor=blue,anchorcolor=blue,citecolor=blue,urlcolor=blue,breaklinks=true]{hyperref}

\def\be{\begin{equation}}
\def\ee{\end{equation}}
\def\bea{\begin{eqnarray}}
\def\eea{\end{eqnarray}}

\begin{document}

\preprint{ADP-22-19/T1190}

\title{Constraints on the dark photon from parity violation and the $W$ mass}

\author{A. W.~Thomas}
\affiliation{ARC Centre of Excellence for Dark Matter Particle Physics and CSSM, Department of Physics, University of Adelaide, Adelaide, SA 5005, Australia}
\date{\today}

\author{X. G.~Wang}
\affiliation{ARC Centre of Excellence for Dark Matter Particle Physics and CSSM, Department of Physics, University of Adelaide, Adelaide, SA 5005, Australia}

\begin{abstract}
We present an analysis of the experimental data for parity-violating electron scattering (PVES) and atomic parity-violation, including the effects of a dark photon.  We derive the favored region of dark photon parameter space, which provides a good description of the experimental data from the Qweak Collaboration and the Jefferson Lab PVDIS Collaboration and simultaneously relieves the tension between the neutron skin thickness determined in the PREX-II experiment and nuclear-model predictions. In addition, we extract the parameter region required to explain the latest W-boson mass anomaly. Our results indicate that a heavy dark photon with mass above the Z boson mass is favored, while other sources of new physics beyond the Standard Model in addition to the dark photon would also be expected.

\end{abstract}

\maketitle

\section{Introduction}

The dark photon as a portal of interactions between dark matter (DM) and Standard Model (SM) particles is one of the promising hypotheses for revealing the nature of DM. 
It has emerged as a canonical scenario, allowing for simple and predictive dark matter models.
We refer to Ref.~\cite{Fabbrichesi:2020wbt} for a review of the current status of theoretical and experimental studies.

A dark photon can be produced in electron or proton fixed-target experiments and at $e^+ e^-$ and hadron colliders, followed by either visible or invisible decays.
Numerous experimental searches have been undertaken~\cite{Merkel:2014avp, LHCb:2019vmc, CMS:2019buh, Banerjee:2019pds, BaBar:2017tiz}.
The strongest limits for $1\ {\rm MeV} \le \bar{m}_{A'} \le 8\ {\rm GeV}$ come from the NA64~\cite{Banerjee:2019pds} experiment and the {\em BABAR} experiment~\cite{BaBar:2017tiz}.
Several planned experiments will explore part of the remaining allowed parameter space~\cite{APEX:2011dww, Battaglieri:2014hga, Beranek:2013yqa}.

Constraints on the dark photon parameter space have also been investigated in connection with the muon $g-2$ anomaly~\cite{Pospelov:2008zw}, 
the electroweak precision observables (EWPO)~\cite{Hook:2010tw, Curtin:2014cca}, and the $e^- p$ deep inelastic scattering (DIS)~\cite{Kribs:2020vyk, Thomas:2021lub, Yan:2022npz}, 
which are independent of the production mechanism and the decay modes of the dark photon.

Recently, we proposed to search for a signal of the effects of a dark photon through parity-violating electron scattering (PVES) experiments and explored the sensitivity of the weak couplings to the dark photon parameters~\cite{Thomas:2022qhj}.
Although the current PVES data are subject to relatively large uncertainties, deviations from SM predictions could imply signals of potential new physics beyond the Standard Model (BSM).
The low energy parity-violation data were applied to constrain new physics in the framework of SM effective field theory (SMEFT)~\cite{Crivellin:2021bkd}.

In this work, we extract the favored region of the dark photon parameter space by fitting the PVES data of the Qweak Collaboration~\cite{Qweak:2018tjf}, 
the PREX-II Collaboration~\cite{PREX:2021umo} and the Jefferson Lab PVDIS Collaboration~\cite{PVDIS:2014cmd}, as well as the most accurate data from atomic parity violation~\cite{Wood:1997zq}. 
We also investigate the implications of this work in the context of the recent W-boson mass anomaly~\cite{CDF:2022hxs}, which has led to many proposed explanations in terms of BSM physics~\cite{Fan:2022dck, Zhu:2022tpr, Athron:2022qpo, Strumia:2022qkt, deBlas:2022hdk, Sakurai:2022hwh, Cheng:2022jyi, Asadi:2022xiy, DiLuzio:2022xns, Gu:2022htv, Cheung:2022zsb, Endo:2022kiw, Nagao:2022oin, Carpenter:2022oyg, Du:2022fqv, Baek:2022agi, Heeck:2022fvl, Yang:2022gvz, Du:2022pbp, Tang:2022pxh, Athron:2022isz, Mondal:2022xdy, Chen:2022ocr, Lu:2022bgw, Song:2022xts, Bahl:2022xzi, Babu:2022pdn, Heo:2022dey, Ahn:2022xeq, Han:2022juu, Arcadi:2022dmt, Ghorbani:2022vtv, Abouabid:2022lpg, Cacciapaglia:2022xih, Biekotter:2022abc, Blennow:2022yfm, Fan:2022yly, Bagnaschi:2022whn, Paul:2022dds, Balkin:2022glu, Zhang:2022nnh, Zeng:2022lkk, Cheng:2022aau, Cai:2022cti, Cao:2022mif, Borah:2022obi, Borah:2022zim, Heckman:2022the, Zhou:2022cql, Yuan:2022cpw, Liu:2022jdq, Kanemura:2022ahw, Popov:2022ldh}. 


\section{Weak couplings}
\label{}
The dark photon was first proposed as an extra $U(1)$ gauge boson~\cite{Fayet:1980ad, Fayet:1980rr, Holdom:1985ag}, 
interacting with SM particles through kinetic mixing with hypercharge~\cite{Okun:1982xi}
\begin{equation}
\mathcal{L}  \supset - \frac{1}{4} F'_{\mu\nu} F'^{\mu\nu} + \frac{m^2_{A'}}{2} A'_{\mu} A'^{\mu} + \frac{\epsilon}{2 \cos\theta_W} F'_{\mu\nu} B^{\mu\nu} \, .
\end{equation}
The physical masses of the Z boson and the dark photon $A_D$ are~\cite{Gopalakrishna:2008dv, Kribs:2020vyk}
\begin{eqnarray}
\label{eq:m_Z_AD}
m^2_{Z, A_D} &=& \frac{m_{\bar{Z}}^2}{2} [ 1 + \epsilon_W^2 + \rho^2 \nonumber\\
&& \pm {\rm sign}(1-\rho^2) \sqrt{(1 + \epsilon_W^2 + \rho^2)^2 - 4 \rho^2} ] \, ,
\end{eqnarray}
where
\begin{eqnarray}
\epsilon_W &=& \frac{\epsilon \tan \theta_W}{\sqrt{1 - \epsilon^2/\cos^2\theta_W}} ,\nonumber\\
\rho &=& \frac{m_{A'}/m_{\bar{Z}}}{\sqrt{1 - \epsilon^2/\cos^2\theta_W}} \, .
\end{eqnarray}
The couplings of the physical $Z$ and dark photon $A_D$ to electrons and quarks are given by~\cite{Gopalakrishna:2008dv, Kribs:2020vyk, Thomas:2022qhj}
 \begin{eqnarray}
\label{eq:C_Z}
C_Z^v &=& (\cos\alpha - \epsilon_W \sin\alpha) C_{\bar{Z}}^v + 2 \epsilon_W \sin\alpha \cos^2 \theta_W C_{\gamma}^v ,\nonumber\\
C_Z^a &=& (\cos\alpha - \epsilon_W \sin\alpha) C_{\bar{Z}}^a ,
\end{eqnarray}
and
\begin{eqnarray}
\label{eq:C_AD}
C_{A_D}^v &=& - (\sin\alpha + \epsilon_W \cos\alpha) C_{\bar{Z}}^v + 2 \epsilon_W \cos\alpha \cos^2 \theta_W C_{\gamma}^v ,\nonumber\\
C_{A_D}^a &=& - (\sin\alpha + \epsilon_W \cos\alpha) C_{\bar{Z}}^a 
\, ,
\end{eqnarray}
respectively, where $\alpha$ is the $\bar{Z}-A'$ mixing angle
\begin{eqnarray}
\tan \alpha &=& \frac{1}{2\epsilon_W} \Big[ 1 - \epsilon^2_W - \rho^2 \nonumber\\
&& - {\rm sign}(1-\rho^2) \sqrt{4\epsilon_W^2 + (1 - \epsilon_W^2 - \rho^2)^2} \Big] \, .
\end{eqnarray}
$C_{\gamma}^v$, $C_{\bar{Z}}^v$, and $C_{\bar{Z}}^a$ are the SM couplings of the photon and the unmixed Z boson,
\begin{eqnarray}
C_{\gamma}^v &=& \{ C^e_{\gamma}, C^u_{\gamma}, C^d_{\gamma}\} = \{ -1, 2/3, - 1/3 \} \, ,\nonumber\\
C_{\bar{Z}}^v &=& \{ g^e_V, g^u_V, g^d_V\} \nonumber\\
                      &=& \{ - \frac{1}{2} + 2 \sin^2\theta_W, \frac{1}{2} - \frac{4}{3}\sin^2\theta_W,
                        \ - \frac{1}{2} + \frac{2}{3}\sin^2\theta_W\} , \nonumber\\
C_{\bar{Z}}^a &=& \{ g^e_A, g^u_A, g^d_A\} = \{ - \frac{1}{2}, \frac{1}{2}, -\frac{1}{2} \} \, .
\end{eqnarray}
By including the $\gamma-Z$ and $\gamma-A_D$ interference contributions to the PVES asymmetry, 
the effective weak couplings at tree level are~\cite{Thomas:2022qhj}
\begin{eqnarray}
\label{eq:C1q-C2q-tree}
C^{\rm tree}_{1q} &=& C^Z_{1q} + \frac{Q^2 + M_Z^2}{Q^2 + M_{A_D}^2} C^{A_D}_{1q} \, ,\nonumber\\
C^{\rm tree}_{2q} &=& C^Z_{2q} + \frac{Q^2 + M_Z^2}{Q^2 + M_{A_D}^2} C^{A_D}_{2q} \,  ,
\end{eqnarray}
where $C^{Z(A_D)}_{1q}$ and  $C^{Z(A_D)}_{2q}$ are the axial-vector vector (AV) and vector axial-vector (VA) combinations of  the electron and the quark couplings, respectively,
\begin{eqnarray}
C^Z_{1q} &=& 2 C_{Z,e}^a C_{Z,q}^v \, ,\ \ C^{A_D}_{1q} = 2 C_{A_D,e}^a C_{A_D,q}^v \, , \nonumber\\
C^Z_{2q} &=& 2 C_{Z,e}^v C_{Z,q}^a\, ,\ \ C^{A_D}_{2q} = 2 C_{A_D,e}^v C_{A_D,q}^a \, .
\end{eqnarray}

Taking into account the radiative corrections~\cite{Erler:2013xha}, the couplings in Eq.~(\ref{eq:C1q-C2q-tree}) will be shifted, 
\begin{eqnarray}
\label{eq:radiative-SM}
C_{1u} & = & \rho_{\Delta} C^{\rm tree}_{1u} - 0.00678 \, ,\nonumber\\
C_{1d} & = & \rho_{\Delta} C^{\rm tree}_{1d} + 0.00078 \, ,\nonumber\\
C_{2u} & = & \rho_{\Delta} C^{\rm tree}_{2u} - 0.01239 \, ,\nonumber\\
C_{2d} & = & \rho_{\Delta} C^{\rm tree}_{2d} + 0.00199 \, ,
\end{eqnarray}
with $\rho_{\Delta} = 1.00064$. In our analysis, we take the SM value $\sin^2\theta_W(\mu = 0)|_{\rm SM} = 0.23857 \pm 0.00005$~\cite{ParticleDataGroup:2020ssz}.


\section{Parity violating data}
The PVES data of the Qweak Collaboration, the PREX-II Collaboration and the Jefferson Lab PVDIS Collaboration are summarized in Tab.~\ref{tab:PVES-DATA}.
 In our analysis, we also include the most accurate atomic parity-violating (APV) data.
 
  %
\begin{table*}[!htpb]
\renewcommand\arraystretch{1.6}
 \begin{center}
\begin{tabular}{ccccc}
\hline\hline
                      {\rm Experiment}                                               &                $Q^2\ ({\rm GeV}^2)$   &                                   {\rm data}                             &        SM             &   SM + dark photon (fit)    \\ \hline
 {\rm Qweak}~\cite{Qweak:2018tjf}                                         &                    $0.0248$                 &            $Q_{\rm w}^p = 0.0719 \pm 0.0045$           &   $0.0708$         &          $0.0707$     \\ 
  {\rm PREX-II}~\cite{PREX:2021umo, Corona:2021yfd}  &                    $0.00616$               & $Q_{\rm w}(^{208}{\rm Pb}) = -114.4 \pm 2.6$        &    $- 117.9$  &       $- 117.1$                     \\
  {\rm PVDIS}~ \cite{PVDIS:2014cmd} $(\times 10^{-6})$      &                     $1.085$                   & $A^{{\rm exp}(1)}_{\rm PV} = -91.1 \pm 3.1 \pm 3.0$    &     $- 87.7$     &        $ - 87.2$        \\
                          \                                                                     &                     $1.901$                  &  $A^{{\rm exp}(2)}_{\rm PV} = -160.8 \pm 6.4 \pm 3.1$  &    $- 158.9$   &        $-157.9$       \\ 
                          APV~\cite{ParticleDataGroup:2020ssz}                             &                          \                         &  $Q_{\rm w}(^{133}{\rm Cs}) = - 72.82(42)$ &    $-73.23$   &   $-72.77$     \\ \hline                  
\end{tabular}
\caption{The PVES and APV data. The fit results including the dark photon effects are given in the last column. }
\label{tab:PVES-DATA}
\end{center}
\end{table*}
%

\subsection{Qweak}
The weak charge of the proton, which characterizes the strength of the proton's interaction with other particles via the neutral electroweak force, is defined in terms of the weak couplings,
\begin{equation}
Q_{\rm w}^p = - 2 ( 2 C_{1u} + C_{1d} )\, .
\end{equation}
The Qweak Collaboration~\cite{Qweak:2018tjf} reported the value
\begin{equation}
Q_{\rm w}^p = 0.0719 \pm 0.0045 \, ,
\end{equation}
derived from the measured parity-violating asymmetry in the scattering of polarized electrons on protons at $Q^2 = 0.0248\ {\rm GeV}^2$, 
which is $-226.5 \pm 9.3$ parts per billion.

\subsection{PREX-II}
The PREX-II Collaboration~\cite{PREX:2021umo} measured the parity-violating asymmetry in the elastic scattering of longitudinally polarized electrons from $^{208}$Pb with very low momentum transfer $Q^2 = 0.00616\ {\rm GeV}^2$.
Using the nuclear weak charge $Q_{\rm w}(^{208}{\rm Pb}) = -117.9$, a neutron skin thickness was extracted 
\begin{equation}
R_{n-p} = R_{n} - R_{p} = 0.283 \pm 0.071\ {\rm fm} \, ,
\end{equation}
which is in mild tension with theoretical nuclear-model predictions~\cite{Piekarewicz:2012pp, Roca-Maza:2015eza, Martinez:2018xep, Piekarewicz:2021jte}, $R_{n-p} = 0.16 - 0.19\ {\rm fm}$, 
and the first {\em ab-initio} estimate, $R_{n-p} = 0.14 - 0.20\ {\rm fm}$~\cite{Hu:2021trw}.

Most recently, it was shown that this tension might be relieved by a small change in the Weinberg angle~\cite{Corona:2021yfd}.
Relying on the theoretical prediction of the neutron skin thickness, $R_{n-p} = 0.16 \pm 0.03\ {\rm fm}$,  and using it as a prior, 
the fitted $\sin^2\theta_W(\mu = 0) = 0.228 \pm 0.008$~\cite{Corona:2021yfd} is about $1.2\sigma$ smaller than the SM prediction.
Using the simple relationship $\delta Q_{\rm w} \approx - 4 Z \delta(\sin^2\theta_W)$, this value implies smaller nuclear weak charge in magnitude of $^{208}$Pb,
\begin{equation}
\label{eq:C_1u-PREX}
Q_{\rm w}(^{208}{\rm Pb}) = -114.4 \pm 2.6\, .
\end{equation}
The weak charge of a nucleus with $Z$ protons and $N$ neutrons can be written in terms of the weak couplings as
\begin{equation}
\label{eq:Qw-Z-N}
Q_{\rm w}^{Z,N} = - 2 (1-\frac{\alpha_{em}}{2\pi}) \Big[ (2 C_{1u} + C_{1d}) Z + (C_{1u} + 2 C_{1d}) N \Big] \, ,
\end{equation}
where $\alpha_{em}$ is the fine structure constant.

In our analysis, we will fit Eq.~(\ref{eq:C_1u-PREX}) instead of the original PREX-II data.

\subsection{PVDIS Collaboration}
The PVDIS Collaboration~\cite{PVDIS:2014cmd} measured the parity-violating asymmetry of electron DIS from deuterium target at two kinematic settings.

The asymmetry at $\langle Q^2 \rangle = 1.085\ {\rm GeV}^2$, $\langle x \rangle = 0.241$, $Y_1 = 1.0$, and $Y_3 = 0.44$ is 
\begin{equation}
A^{{\rm exp}(1)}_{\rm PV} = [ -91.1 \pm 3.1 (\rm stat.) \pm 3.0 (\rm syst.) ] \times 10^{-6} \, .
\end{equation}
The second DIS setting is $\langle Q^2 \rangle = 1.901\ {\rm GeV}^2$, $\langle x \rangle = 0.295$, $Y_1 = 1.0$, and $Y_3 = 0.69$, with the result
\begin{equation}
A^{{\rm exp}(2)}_{\rm PV} = [- 160.8 \pm 6.4 (\rm stat.) \pm 3.1(\rm syst.) ] \times 10^{-6} \, .
\end{equation}
The sensitivities of these asymmetries to the weak couplings were given by~\cite{PVDIS:2014cmd}
\begin{eqnarray}
A^{(1)}_{\rm PV} &=& ( 1.156\times 10^{-4} ) [ (2 C_{1u} - C_{1d}) + 0.348 (2 C_{2u} - C_{2d})] \, , \nonumber\\
A^{(2)}_{\rm PV} &=& ( 2.022 \times 10^{-4} ) [ (2 C_{1u} - C_{1d}) + 0.594 (2 C_{2u} - C_{2d})] \, , \nonumber\\
\end{eqnarray}
respectively, which were calculated using the MSTW2008 leading-order parametrization of parton distribution functions (PDF)~\cite{Martin:2009iq}.

For this experiment, effects arising because $Q^2 \ne 0$ shift the weak couplings in Eq.~(\ref{eq:radiative-SM}) by~\cite{Erler:2013xha}, 
\begin{eqnarray}
\label{eq:C_1q}
C_{1u} & \rightarrow & C_{1u} - \frac{0.0022}{3} \ln\frac{Q^2}{0.14\ {\rm GeV}^2}\, ,\nonumber\\
C_{1d} & \rightarrow & C_{1d} + \frac{0.0011}{3} \ln\frac{Q^2}{0.14\ {\rm GeV}^2}
\, ,
\end{eqnarray}
and
\begin{eqnarray}
\label{eq:C_2q}
C_{2u} & \rightarrow & C_{2u} - 0.0009 \ln\frac{Q^2}{0.078\ {\rm GeV}^2}\, ,\nonumber\\
C_{2d} & \rightarrow & C_{2d} + 0.0007 \ln\frac{Q^2}{0.021\ {\rm GeV}^2}\, .
\end{eqnarray}
%

\subsection{Atomic parity-violation data}
The very accurate measurement of the parity non-conserving (PNC) $6s-7s$ amplitude in cesium~\cite{Wood:1997zq}, $E_{PNC}$, has been widely applied in searches for new physics beyond the SM. In combination with the vector transition polarizability~\cite{Bennett:1999pd}, $\beta$, the weak charge of $^{133}$Cs was extracted using state-of-the-art many body calculations~\cite{Porsev:2009pr}, with a result that was in perfect agreement with the SM prediction.
However, a later analysis including significant corrections to two non-dominating terms yielded~\cite{Dzuba:2012kx}  
\begin{equation}
\label{eq:Qw-Cs}
Q_{\rm w}(^{133}{\rm Cs}) = - 72.58(29)_{\rm expt} (32)_{\rm theo}\, ,
\end{equation}
which deviates from the SM value by $1.5\sigma$. 
Most recently, it was shown that this discrepancy could be reduced by incorporating the latest W-boson mass anomaly through a change in $\sin^2\theta_W$~\cite{Tan:2022bip}.

Other determinations led to slightly different results~\cite{Cadeddu:2021dqx, Sahoo:2021thl, Roberts:2021esp}.
The latest Particle Data Group (PDG) value is~\cite{ParticleDataGroup:2020ssz}
\begin{equation}
\label{eq:Qw-Cs-PDG}
Q_{\rm w}(^{133}{\rm Cs})_{\rm PDG} = - 72.82(42)\, .
\end{equation}

In our analysis, we fit Eq.~(\ref{eq:Qw-Cs-PDG}) in the dark photon framework through Eq.~(\ref{eq:Qw-Z-N}), while retaining the SM value of $\sin^2\theta_W$.

\section{Results}
\label{sec:results}
\begin{figure}[!h]
\includegraphics[width=\columnwidth]{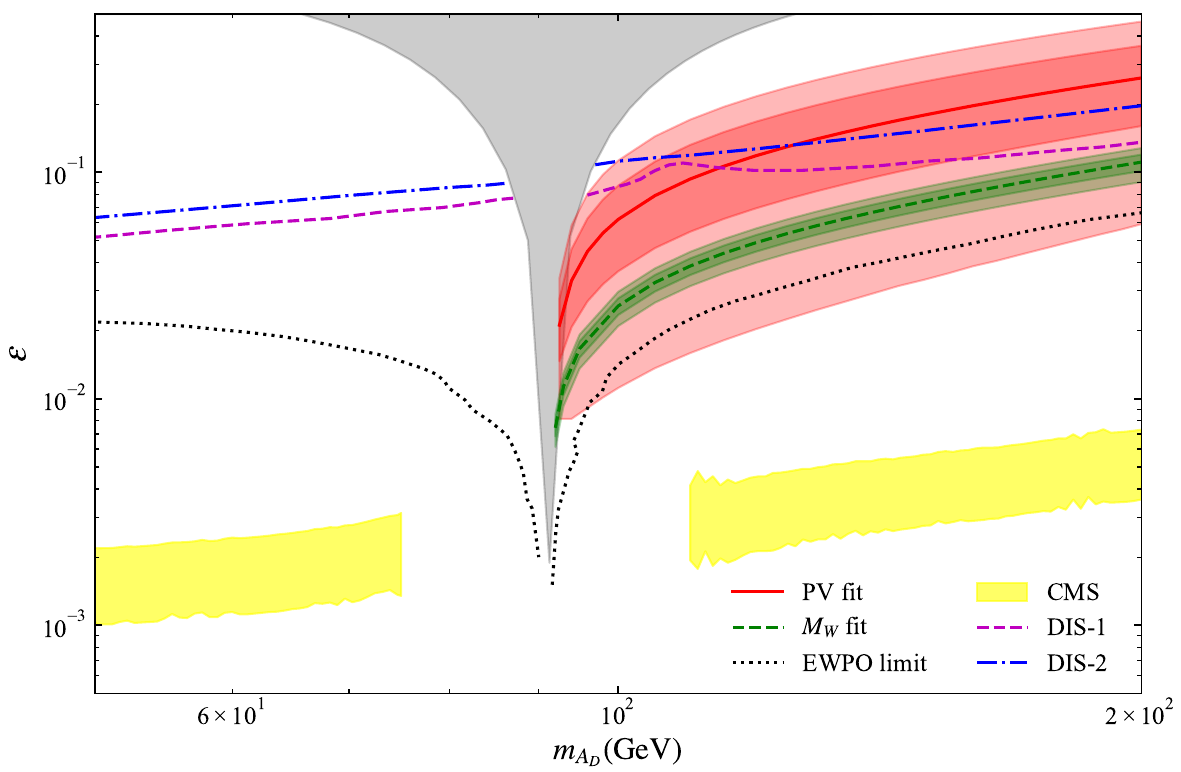}
\vspace*{-0.1cm}
\caption{The best values of $\epsilon$ obtained by fitting the parity-violating data (red solid) and the W-boson mass (green dashed), together with 68\% CL (dark band) and 95\% CL (light band) uncertainties.
The region in grey is not accessible because of the ``eigenmass repulsion" associated with the $Z$ mass.
The EWPO and DIS-1 limits are taken from Refs.~\cite{Curtin:2014cca} and \cite{Kribs:2020vyk}, respectively. 
The DIS-2 constraint is taken from Ref.~\cite{Thomas:2021lub} (95\% CL) by extending the parameter region up to $M_{A_D} = 200\ {\rm GeV}$. 
We also show the 95\% CL exclusion limits from the CMS Collaboration~\cite{CMS:2019buh}.}
\label{fig:best-epsilon-PDG}
\end{figure}

We perform a $\chi^2$ fit to the parity-violating (PV) data in Tab.~\ref{tab:PVES-DATA}.
For each value of $M_{A_D}$, the best values of $\epsilon$ are shown in Fig.~\ref{fig:best-epsilon-PDG} with $\chi^2_{\rm total} = 2.179$, compared with the value $\chi^2_{\rm total} = 3.517$ without the dark photon.
The results favor the region with $M_{A_D} > M_Z$, where the dark photon corrections to $C_{1q}$ and $C_{2q}$ at low $Q^2$ are negative ~\cite{Thomas:2022qhj}.
The region with $M_{A_D}$ below the Z boson mass is disfavored. 
The corresponding fit results are given in the last column of Tab.~\ref{tab:PVES-DATA}.
Our analysis also provides a possible resolution to relieve the tension between the PREX-II and nuclear-model predictions without changes in the weak mixing angle.

In Fig.~\ref{fig:best-epsilon-PDG}, we also show the existing exclusion limits from decay-agnostic processes. 
The EWPO determination~\cite{Curtin:2014cca} tends to place strong constraints on the dark photon parameters, 
while the upper limits on $\epsilon$ from the $e^- p$ DIS process~\cite{Kribs:2020vyk, Thomas:2021lub} are relatively weaker, especially in the heavy mass region of the dark photon.

However, one of the physical inputs in EWPO analysis is the mass of the W-boson.  Recently, the Collider Detector at Fermilab (CDF) reported the most accurate measurement of the W-boson mass~\cite{CDF:2022hxs},
\begin{equation}
\label{eq:m_W}
m_W = 80.4335 \pm 0.0094\ {\rm GeV} \, ,
\end{equation}
which constitutes a $7\sigma$ deviation from the SM prediction.
The current EWPO constraint~\cite{Curtin:2014cca} could be weakened by taking this anomaly into account.
Quantitative analysis requires a global fit to all available electroweak observables.

The relation between the W-boson mass, $m_W$, and the the Z-boson mass $m_{\bar{Z}}$ is~\cite{Awramik:2003rn},
\begin{equation}
m^2_W = m^2_{\bar Z} \left\{ \frac{1}{2} + \sqrt{\frac{1}{4} - \frac{\pi \alpha_{em}}{\sqrt{2} G_F m^2_{\bar Z}} [1 + \Delta r(m_W, m_{\bar Z}, m_H, m_t,\ldots)]} \right\} \, ,
\end{equation}
where $G_F = 1.1663787 \times 10^{-5}\ {\rm GeV}^{-2}$ is the Fermi constant, and
\begin{eqnarray}
\Delta r &=& \Delta r^{(\alpha)} + \Delta r^{(\alpha \alpha_s)} + \Delta r^{(\alpha \alpha_s^2)} + \Delta r^{(\alpha^2)}_{\rm ferm} + \Delta r^{(\alpha^2)}_{\rm bos} \nonumber\\
&&  + \Delta r^{(G_F^2 \alpha_s m_t^4)} + \Delta r^{(G_F^3 m_t^6)} .
\end{eqnarray}
Using $m_H = 125.14\ {\rm GeV}$ and $m_t = 172.89\ {\rm GeV}$, we derive $\Delta r = 0.03677$.
Then Eq.~(\ref{eq:m_W}) implies $m_{\bar Z} = 91.2326 \pm 0.0076\ {\rm GeV}$.

In Fig.~\ref{fig:best-epsilon-PDG}, we also show the region of dark photon parameters (green band) which are required to shift $m_{\bar Z}$ to the physical value $m_Z = 91.1875 \pm 0.0021\ {\rm GeV}$ through Eq.~(\ref{eq:m_Z_AD}).
Similar results were obtained by calculating the oblique parameters $S$, $T$ and $U$ in the dark photon framework~\cite{Zeng:2022lkk, Cheng:2022aau}.

Both the PV fit and the W-boson anomaly favor a heavy dark photon with mass above $m_Z$.
In that region these two constraints are compatible within the estimated uncertainties, while they are in serious tension with the most stringent constraints from the CMS Collaboration~\cite{CMS:2019buh} for $m_{A_D}$ in the range 110-200 GeV. 
The region of parameter space with $m_Z < m_{A_D} < 110\ {\rm GeV}$, which is unconstrained by the CMS limits, as the Z boson dominates $\mu^+ \, \mu^-$ production there, seems to be the most promising region to simultaneously describe the parity-violating data and the W-boson mass.
%

\section{Conclusion}
We have investigated the possibility of a simultaneous fit to the experimental data involving parity-violating electron scattering and atomic parity-violation within the framework of physics beyond the Standard Model arising from a dark photon. 
We extracted the favored region of dark photon parameter space, which is required to reach a good description of these PVES and APV data.
It also provides a possible resolution to relieve the tension of neutron skin thickness between PREX-II determination and nuclear-model predictions.
We also explored the region of dark photon parameters required to explain the latest W-boson mass anomaly from the CDF Collaboration. 

Both the fits to parity-violating data and the W-boson mass favor a heavy dark photon with mass above the Z-boson mass.
These two constraints are compatible within the estimated uncertainties.
With a dark photon as the only new physics beyond the Standard Model, the favored parameter space from the parity-violation data and the W mass appear incompatible with the constraints from the CMS Collaboration for a dark photon mass between 110 and 200 GeV. On the other hand, there remains an important region of dark photon mass between the $Z$-boson mass and 110 GeV, where there is no other constraint in tension with the dark photon explanation.

We expect that the proposed PVES measurements at SoLID~\cite{Chen:2014psa}, P2~\cite{Becker:2018ggl} and MOLLER~\cite{MOLLER:2014iki} should provide more stringent constraints on the dark photon parameters. 

\section*{Acknowledgments}
We thank Xiaochao Zheng, Jens Erler, and Ross Young for helpful discussions.
This work was supported by the University of Adelaide and the Australian Research Council through the Centre of Excellence for Dark Matter Particle Physics (CE200100008).


\end{document}